\documentclass[nofootinbib, twocolumn,
		        aps,
		        superscriptaddress,
		        pra
		       ]{revtex4-1}
		
\usepackage[colorlinks=true,urlcolor=blue,citecolor=blue,linkcolor=red]{hyperref}
\usepackage{graphicx}
\usepackage{wrapfig}
\usepackage{booktabs}
\usepackage{amsmath}
\usepackage{amssymb}
\usepackage{braket}
\usepackage[dvipsnames]{xcolor}
\usepackage[normalem]{ulem}
\usepackage{siunitx}
\usepackage[T1]{fontenc}
\usepackage{hyperref}
\usepackage{bbold}

\begin{document}
	\title{Relativistic limits on the discretization and temporal resolution of a quantum clock}

	\author{Tommaso Favalli}\email{tommaso.favalli@units.it}
	\affiliation{University of Trieste, Strada Costiera 11, I-34151 Trieste, Italy}

	\begin{abstract}
We provide a brief discussion regarding relativistic limits on the discretization and temporal resolution of time values in a quantum clock. Our clock is characterized by a time observable chosen to be the complement of a bounded and discrete Hamiltonian which can have an equally-spaced or a generic spectrum. In the first case the time observable can be described by a Hermitian operator and we find a limit in the discretization for the time eigenvalues. Nevertheless, in both cases, the time observable can be described by a POVM and, by increasing the number of time states, we show how the bound on the minimum time quantum can be reduced and identify the conditions under which the clock values can be treated as continuous. Finally, we find a limit for temporal resolution of our time observable when the clock is used (together with light signals) in a relativistic framework for measuring spacetime distances.
	\end{abstract}
	
	\maketitle
	
\section{Introduction}
It is well known that fundamental limits emerge in the precision with which to measure space and time when quantum mechanics and the theory of relativity are considered together \cite{gambinilimit,limit1,limit2,limit3,limit4} (see also \cite{greinerrelativistic,peskin,librorovelli}). 
Our purpose in this work is to study the relativistic limits in the discretization and temporal resolution of time values in a quantum clock. 
A good definition of a clock can be found in the work by Peres \cite{peres}:<<A clock is a dynamical system which passes through a succession of states at constant time intervals>>.
Our quantum clock is described by a quantum time observable chosen to be the complement of a bounded and discrete Hamiltonian which can have both an equally-spaced or a generic spectrum. Only in the case of equally-spaced energy spectrum the time observable can be described by a Hermitian operator while, in general, it will be described by a POVM.
This kind of observable was introduced by Pegg in \cite{pegg} (see also \cite{chapter3,opfase,peggbar,jan1,pereslibro}). It generalizes the quantum clock proposed by Salecker and Wigner (SW) in \cite{wignerlimits,wignerlimit}, where the authors used clocks and light signals in order to measure distances between spacetime events. 

Pegg's clock is, for example, widely used as a possible choice of time observable in the Page and Wootters quantum time formalism \cite{pagewootters,wootters} (see also \cite{librotommi,nostro,nostro2,nostro3,nostro4,lloydmaccone,vedral,vedraltemperature,macconeoptempoarrivo,interacting,timedilation,simile,simile2,esp1,esp2} and references therein). However, we do not adopt this framework here, and we treat our time states (to be introduced) as evolving in an \textit{external time} parameter, which is standard in quantum mechanics. 
We will return to the connection with the Page and Wootters approach in the Conclusions.

As we will see in the following, Pegg's observable can exhibit both discrete and continuous time values, and we therefore ask \textit{whether there is a fundamental limit in the spacing between discrete time values of clock and whether time values can be safely considered as continuous}. Finally, although the time values may be continuous, we ask \textit{what is the minimum resolvable interval between them} when the clock operates within the relativistic SW framework, that is, what is the temporal resolution $\Delta t$ (in what follows, we will refer to $\Delta t$ also as temporal accuracy). Notice that throughout the text we use \textit{time values} or \textit{clock values} to refer to the values of the clock observable, and not to the external time parameter.

We perform our investigation first considering the clock with equally-spaced energy spectrum in Section II; then we generalize the discussion for a clock with generic spectrum in Section III. 
We emphasize that, since any realistic quantum clock is a system with finite size, the introduction of unbounded Hamiltonians with continuous spectrum would not be possible. This is the reason why we choose to focus on bounded and discrete Hamiltonians in describing our quantum clock, also considering that this may encourage experimental applications.


\section{Clock with equally-spaced energy spectrum} 
\subsection{The quantum clock}
We introduce the clock in the case of equally-spaced energy spectrum.
The non-degenerate clock Hamiltonian can be written as:
\begin{equation}\label{Hclock}
	\hat{H} = \sum_{n=0}^{p} E_n \ket{E_n}\bra{E_n}
\end{equation}
where
\begin{equation}\label{eigenclock}
	E_n= \frac{2\pi \hslash}{T}n 
\end{equation}
with $n=0,1,...,p$ and $p+1 = d$ dimension of the Hilbert space $\mathcal{H}$ of our clock system. The meaning of $T$ will become clear soon. 
Notice that we have chosen $E_0 = 0$ to avoid unnecessary complications. The case with $E_0 \neq 0$ can be considered as a generalization, which we do not address here.
Next, we introduce the time observable by defining $z+1\ge p+1 = d$ time states:
\begin{equation}\label{timestates}
	\ket{\tau_m} = \frac{1}{\sqrt{p+1}} \sum_{n=0}^{p} e^{-i \hslash^{-1}E_n \tau_m} \ket{E_n}
\end{equation}
where
\begin{equation}\label{timevalues}
\tau_m = \tau_0 + \frac{T}{z+1} m 
\end{equation}
with $m=0,1,...,z \ge p$. The states (\ref{timestates}) exhibit a cyclic condition 
and the meaning of $T$ is now clear: it represents the time taken by the clock to return to its initial state. We therefore consider three cases of interest: 
\begin{itemize}
	\item for $z=p$ we can introduce the Hermitian operator:
	\begin{equation}\label{optau}
		\hat{\tau} = \sum_{m=0}^{p} \tau_m \ket{ \tau_m}\bra{\tau_m}
	\end{equation}
	where the time states $\ket{\tau_m}$ provide an orthonormal and complete basis in $\mathcal{H}$. 
	The operator (\ref{optau}) can be considered as conjugated (or complement) to the Hamiltonian $\hat{H}$: it is indeed easy to show
	that $\hat{H}$ is the generator of shifts in $\tau_m$ values and, viceversa, $\hat{\tau}$ is the generator of energy shifts \cite{nostro};
	
	\item for $z>p$ the number of time states is greater than the number of energy states and the time observable is represented by a POVM with $z+1$ elements $\frac{p+1}{z+1} \ket{ \tau_m}\bra{\tau_m}$. The resolution of identity $\frac{p+1}{z+1}\sum_{m=0}^{z} \ket{ \tau_m}\bra{\tau_m} = \mathbb{1}$ is indeed still satisfied even if the time states are not orthogonal;
	
	\item in the limiting case $z\rightarrow\infty$ is possible to redefine the time states as
	\begin{equation}\label{timestatecont}
		\ket{t} = \frac{1}{\sqrt{p+1}} \sum_{n=0}^{p} e^{-i\hslash^{-1} E_n t}\ket{E_n}
	\end{equation}
	where $t$ is a continuous variable taking all the real values in the interval $\left[t_0,t_0+T\right]$. The time observable is again described by a POVM generated with the operators $\frac{p+1}{T}\ket{t}\bra{t}dt$ and the resolution of the identity reads here $\frac{p+1}{T}\int dt \ket{t}\bra{t} =\mathbb{1}$.
\end{itemize} 
In the first two cases the time values of the clock are discrete and we can search the fundamental limit in the spacing between them. We will than see that such limit reduces to zero in the case of $z\rightarrow\infty$, allowing us to safely consider the time values as continuous in this limiting case. Finally, we will derive a bound on the minimal resolvable interval in the clock time values.


\subsection{Limit in discretizing time}
We proceed here in our analysis by considering $z \ge p$ and we will study $z=p$ as a special case. From (\ref{eigenclock}) and (\ref{timevalues}) we can easily calculate the spacing between two neighbors time eigenvalues:
\begin{equation}\label{dtau}
	\delta \tau = \tau_{m + 1} - \tau_{m} = \frac{2\pi \hslash}{\delta E \left(z+1\right)} = \frac{T}{z+1}
\end{equation}
where $\delta E = 2\pi\hslash/T$ is the interval between two neighbors energy levels. 
We now search a lower limit for $\delta \tau$.
The extent of the energy spectrum is $\delta E \left(p+1\right)$ and it can be bound considering that \textit{energy cannot be arbitrarily confined in a region of space}. Assuming the clock to be spherically symmetric, we thus require that (half) the diameter of the clock $l_C$ not be smaller than its Schwarzschild radius \cite{limit2,limit3,librorovelli}. 
Denoting as $m_{\text{rest}}$ the rest mass of the clock (excluding its internal energy), we therefore ask:
\begin{equation}\label{8}
	\delta E  \left(p+1\right) + m_{\text{rest}}c^2 < \frac{l_C}{2} \frac{c^4}{2G} 
\end{equation}
which leads to
\begin{equation}
	\delta E (z+1) < \left(\frac{l_C c^4}{4 G} - m_{\text{rest}}c^2 \right)  \frac{(z+1)}{(p+1)}
\end{equation}
and finally to
\begin{equation}\label{chiave}
	\delta \tau > 2 \pi \left(\frac{l_C}{4l_p t_p} - \frac{m_{\text{rest}}c^2}{\hslash} \right)^{-1} \frac{(p+1)}{(z+1)}
\end{equation}
where we have introduced the Planck length and the Planck time as 
\begin{equation}
	l_p= \sqrt{\hslash G/c^3} \quad \text{and} \quad t_p =  \sqrt{\hslash G/c^5} .
\end{equation}
It is worth noting that the condition $l_C/2 > 2Gm_{\text{rest}}/c^2$ guarantees that the expression $ \left(\frac{l_C}{4l_p t_p} - \frac{m_{\text{rest}}c^2}{\hslash} \right)$ (appearing in the denominator of equation~(\ref{chiave})) remains positive.

In the case $z=p$, where the time states are orthogonal and we can introduce the Hermitian operator $\hat{\tau}$, equation (\ref{chiave}) becomes 
\begin{equation}\label{11}
	\delta \tau_{z=p} > 2 \pi \left(\frac{l_C}{4l_p t_p} - \frac{m_{\text{rest}}c^2}{\hslash} \right)^{-1}
\end{equation}
which imposes a bound in the minimum spacing between time values depending on $m_{\text{rest}}$ and $l_C$.

For $z>p$, we can instead obtain an arbitrarily small bound for $\delta \tau$ by taking $z$ arbitrarily large. Furthermore, under the condition $T>2 \pi (p+1)/\left(\frac{l_C}{4l_p t_p} - \frac{m_{\text{rest}}c^2}{\hslash} \right)$
(which is satisfied in any realistic scenario), equations (\ref{chiave}) and (\ref{dtau}) ensure that we can safely take the limit $z\rightarrow\infty$, where time values become continuous. Indeed, as $\delta \tau \to 0$, also the constraint on $\delta \tau$ tends to zero (with the same scaling) and $\delta \tau$ always remains above the bound.


\subsection{Limit in resolving time}
We work here in the limit $z\rightarrow\infty$, where the time states are defined as in (\ref{timestatecont}) and the continuous variable $t$ takes all the real values in the interval $\left[t_0,t_0+T\right]$. The eigenvalues of the clock Hamiltonian $\hat{H}$ does not change and they are again described by (\ref{eigenclock}).
Our goal is to identify a fundamental limit for the minimal resolvable interval between time values when relativistic considerations are taken into account. 
In a truly relativistic framework the concept of time is closely connected with the concept of space. 
The basic measurement in General Relativity is indeed the measurement of distances between events in spacetime. Such measurements make the definition of a coordinate system possible.
For this reason, in deriving the bound for the temporal accuracy, we consider the SW framework in which clocks and light signals are used to measure spacetime distances \cite{wignerlimits,wignerlimit}. In some sense, we are doing the opposite of what SW did: we are not looking for quantum limitations to General Relativity, but instead we are searching the limitations that a relativistic framework imposes on quantum theory.

The distances between events in spacetime could be measured using clocks and rods but, as Wigner observes, <<we found that measurements with yardsticks are rather difficult to describe and that their use would involve a great deal of unnecessary complications
[...] It is desirable, therefore, to reduce all measurements in spacetime to measurements by clocks>> \cite{wignerlimit}. Clearly, only time-like distances between events can be measured by clocks directly, while space-like distances between events (which would naturally be measured by rods) have to be measured indirectly with the help of light signals.   

Summarizing SW's argument: the simplest framework in spacetime capable of measuring distances between events is a set of clocks 
together with light signals. The clocks 
should be only slowly moving with respect to each other, namely with world lines approximately parallel\footnote{For a detailed discussion regarding spacetime measurements between events with clocks we directly refer to \cite{wignerlimits,wignerlimit}.}.
 
We thus consider our clock with accuracy $\Delta t$, being able to measure time intervals up to a maximum $T$. 
The~time state of the clock is given by equation (\ref{timestatecont}), which we rewrite to facilitate the reading: 
\begin{equation}
		\ket{t} =\frac{1}{\sqrt{p+1}} \sum_{n=0}^{p} e^{-i\hslash^{-1} E_n t}\ket{E_n} .
\end{equation}
Such state evolves through $p+1=d$ orthogonal states at time values $\frac{kT}{p+1}$ with $k=0,1,2,...,p$, implying a \textit{first bound} (which we will call the \textit{structural bound}) on the temporal accuracy $\Delta t$: 
\begin{equation}
	\frac{T}{\Delta t }\le p+1 
\end{equation}
leading to
\begin{equation}\label{15}
	\Delta t \ge \frac{T}{p+1} = \delta \tau_{z=p} .
\end{equation}
Notice that we already found the lower limit for $ \delta \tau_{z=p}$ in equation (\ref{11}).



We now move to the \textit{second bound} for the clock. Assuming the clock as not absolutely stationary, 
we can think of it as having an indeterminate momentum and thus a spread in the velocity.
From the Heisemberg uncertainty principle, identifying with $\delta x$ the spread in the position, we have that such spread in the velocity is\footnote{We notice and emphasize here that we are doing a strong semplification restricting the spreads in position and velocity only to one spatial dimension. For a detailed discussion about physical arguments underlying this assumption we refer to \cite{wignerlimits}.}:
\begin{equation}
	\delta v = \frac{\delta p }{m} \sim \frac{\hslash}{2m\delta x}
\end{equation}
where $m$ denotes the inertial (and gravitational) mass of the clock (with $m \approx m_{\text{rest}}$ in any realistic setting). Therefore, after a time interval $\Theta\le T$ (denoting the \textit{operational time}), the uncertainty in the position becomes $\delta x + \frac{\hslash  \Theta}{2m\delta x}$, which (given the mass $m$) assumes its minimum for: $\delta x = \left(\frac{\hslash  \Theta}{2m}\right)^{\frac{1}{2}}$. 
We now ask that the clock position does not introduce any statistical uncertainty in the determination of time. That is, we assume that the position spread is small enough so that the uncertainty in the time at which the clock interacts with a reference event (i.e. a light signal) remains within a time interval $\Delta t$.
It can be done by requiring $\delta x \lesssim c \Delta t$ throughout the interval $\Theta$, namely
\begin{equation}\label{xmin}
 \delta x =
  \left(\frac{\hslash  \Theta}{2m}\right)^{\frac{1}{2}} \lesssim c \Delta t .
\end{equation}
On the other side, to ensure gravitational consistency, we ask that the spatial uncertainty of the clock be no smaller than (twice) its Schwarzschild radius:
\begin{equation}\label{xmax}
2\frac{2Gm}{c^2}	<  \delta x = \left(\frac{\hslash  \Theta}{2m}\right)^{\frac{1}{2}} 
\end{equation}
which leads to the requirement for the mass
\begin{equation}\label{massreq}
	m < \left( \frac{c^4 \hslash \Theta}{32 G^2}\right)^{\frac{1}{3}}.
\end{equation}
Combining now together equations (\ref{xmin}) and (\ref{massreq}) we obtain
\begin{equation}
	c^2 \Delta t^2 \gtrsim \frac{\hslash \Theta}{2 m}  > \frac{\hslash \Theta}{2}  \left( \frac{32 G^2}{c^4 \hslash \Theta}\right)^{\frac{1}{3}} 
\end{equation}
and finally:
\begin{equation}\label{limitealtri}
\Delta t > 2^{\frac{1}{3}} \Theta^{\frac{1}{3}}t^{\frac{2}{3}}_p .
\end{equation} 
Our derivation combines the quantum-spreading argument of SW \cite{wignerlimits,wignerlimit} with a gravitational consistency condition requiring that the clock cannot be localized within its own Schwarzschild radius. This leads to a bound on the time resolution of our clock, aligning with previous analyses that have explored fundamental limits on time measurement precision (see \cite{limit1,limit2,limit3,limit4}). In particular, we highlight the findings of Gambini and Pullin \cite{gambinilimit}, who derive a similar limit based on time dilation effects. 


Importantly, we have verified that including the Compton wavelength bound 
\begin{equation}
	\frac{\hslash}{mc}   \lesssim \delta x \lesssim c \Delta t
\end{equation}
together with the spreading constraint (\ref{xmin}) and the gravitational consistency condition 
does not affect our result when $\Theta > 4 t_p$. 
Indeed, in this physically relevant regime,
minimizing $\Delta t$ over the clock mass $m$ subject to all the three bounds yields an optimal mass 
for which the spreading and gravitational constraints coincide and jointly determine the final limit, 
while the Compton bound is less restrictive and plays no role. 
For sufficiently small clock masses, the Compton bound would dominate. However, in such cases it lies above the other two bounds and therefore does not lower the achievable $\Delta t$. 

 For completeness, we note that for $\Theta < 4 t_p$ the optimal clock mass obtained by minimizing $\Delta t$ under the three bounds is determined by the intersection of the Compton and gravitational constraints, which sets a lower floor $\Delta t \gtrsim 2 t_p$ (this same floor also occurs at $\Theta \approx 4 t_p$, where the optimal mass lies at the intersection of all three bounds). We stress, however, that all physically realistic scenarios correspond to $\Theta \gg t_p$, where (\ref{limitealtri}) applies. 

We emphasize that, in realistic parameter regimes such as those corresponding to currently achievable clock spectra, the structural bound $\Delta t \ge T /(p + 1)$ dominates over our second fundamental constraint (\ref{limitealtri}). Moreover, we note that, for physically accessible clock masses (typically much smaller than the mass scale optimizing the second bound), the quantum--spreading constraint $c \Delta t \gtrsim \left(\frac{\hbar T}{2m}\right)^{\frac{1}{2}}$ becomes the relevant competitor as second bound, so the structural one must be compared with it to determine the actual limitation on clock accuracy. Nevertheless, equation~(\ref{limitealtri}) retains its role as the ultimate fundamental limitation that cannot be surpassed, even if it lies beyond the range of any physically realizable implementation.


\section{Clock with generic spectrum}
\subsection{The generalized quantum clock}
We consider here again the clock as a quantum system described by $d=p+1$ energy states $\ket{E_n}$ and $E_n$ energy levels with $n=0,1,2,...,p$, but we do not assume an equally-spaced energy spectrum. In this case we cannot find a subset of $p+1$ time states (\ref{timestates}) that are orthogonal but we can make progress by requiring that the ratios $E_n / E_1$ are rational numbers. Thus we can write: 
\begin{equation}\label{3numraz1} 
	\frac{E_n}{E_1} = \frac{C_n}{B_n} 
\end{equation}
where $C_n$ and $B_n$ are integers with no common factors. We define $r_n = r_1 C_n/B_n$ for $n>1$, with $r_1$ the lowest common multiple of the values of $B_n$ with $n>1$, and we take $r_0=0$. In this framework the values $r_n$ are integers for all $n \ge 0$. Now we redefine
\begin{equation}\label{defT2}
	T=\frac{2\pi \hslash r_1}{E_1}
\end{equation} 
and then 
\begin{equation}\label{en2}
	E_n = r_n \frac{2\pi \hslash}{T}.
\end{equation} 

In this framework, we introduce again the $z+1$ states:
\begin{equation}\label{timestates2}
	\ket{\tau_m} = \frac{1}{\sqrt{p+1}} \sum_{n=0}^{p} e^{-i \hslash^{-1} E_n \tau_m} \ket{E_n}
\end{equation}
with $\tau_m = \tau_0 + \frac{T}{z+1} m $.
These states still satisfy the resolution of the identity, indeed we have:
\begin{widetext}
	\begin{equation}\label{311111}
	\begin{split}
		\sum_{m=0}^{z}\ket{\tau_m}\bra{\tau_m} & = \frac{1}{p+1} \left \{ \sum_{m=0}^{z}\sum_{n=0}^{p}\sum_{k=0}^{p} e^{-i \hslash^{-1}  (E_n - E_k) \tau_m} \ket{E_n}\bra{E_k} \right \} 
		 \\& = \frac{1}{p+1} \left \{ \sum_{m=0}^{z}\sum_{n=0}^{p}\ket{E_n}\bra{E_n}  + \sum_{m=0}^{z} \sum_{n \ne k} e^{-i \frac{2\pi}{T} (r_n - r_k) \tau_m } \ket{E_n}\bra{E_k} \right \}.
	\end{split}
\end{equation} 
Replacing the expression of $\tau_m$ in the second term on the right-hand side of the equation (\ref{311111}), we obtain:
\begin{equation}\label{32222}
	\begin{split}
		\sum_{m=0}^{z}\ket{\tau_m}\bra{\tau_m} & = \frac{1}{p+1} \left \{ \sum_{m=0}^{z}\sum_{n=0}^{p}\ket{E_n}\bra{E_n}  + \sum_{m=0}^{z} \sum_{n \ne k} e^{-i \frac{2\pi}{T}(r_n - r_k) (\tau_0 + m \frac{T}{z+1}) } \ket{E_n}\bra{E_k} \right \} 
		\\& =  \frac{1}{p+1} \left \{ \sum_{m=0}^{z}\sum_{n=0}^{p}\ket{E_n}\bra{E_n}  + \sum_{n \ne k} e^{i  \frac{2\pi}{T}(r_n - r_k)\tau_0} \sum_{m=0}^{z} e^{-i  \frac{2 \pi m}{z+1}(r_n - r_k)} \ket{E_n}\bra{E_k} \right \}.
	\end{split}
\end{equation} 
\end{widetext}
For $E_n / E_1$ rational, and thus $r_n - r_k$ an integer, we have 
\begin{equation}
	\sum_{n \ne k} e^{i  \frac{2\pi}{T}(r_n - r_k)\tau_0} \sum_{m=0}^{z} e^{-i  \frac{2 \pi m}{z+1}(r_n - r_k)} \ket{E_n}\bra{E_k} = 0 
\end{equation}
by virtue of $\sum_{m=0}^{z} e^{-i  \frac{2 \pi m}{z+1}(r_n - r_k)} = (z+1)\delta_{n,k}$, which ensures the cancellation of all terms with $n \neq k$.
Combining all the above, we thus finally obtain:
\begin{equation}\label{33333}
	\frac{p+1}{z+1}\sum_{m=0}^{z}\ket{\tau_m}\bra{\tau_m} = \mathbb{1} .
\end{equation} 
We can ensure $r_n - r_k$ is not a multiple of $z+1$ by taking $z+1 > r_p$, that is the largest value for $r_n$. This implies that, in this new scenario, the generalized quantum clock is only described by the POVM, where the $z+1$ non-orthogonal elements are given by $\frac{p+1}{z+1} \ket{ \tau_m}\bra{\tau_m}$.

As in the previous Section, since $z$ is lower-bounded, we can take the limit $z\rightarrow \infty$, defining the time states as
\begin{equation}\label{timestatecont2}
	\ket{t} = \frac{1}{\sqrt{p+1}} \sum_{n=0}^{p} e^{-i\hslash^{-1} E_n t}\ket{E_n}
\end{equation}
where $t\in\left[t_0,t_0+T\right]$. The clock is here described by the POVM generated with the operators $\frac{p+1}{T}\ket{t}\bra{t}dt$ and the resolution of the identity reads 
\begin{equation}\label{33334}
	\frac{p+1}{T}\int_{t_0}^{t_0+T} dt \ket{t}\bra{t} =\mathbb{1}.
\end{equation}
To conclude the paragraph we emphasize that this framework allow us to use any generic (discrete) clock Hamiltonian with arbitrary (not rational) energy level ratios. In this case, the resolutions of the identity (\ref{33333}) and (\ref{33334}) are no longer exact and the time states do not provide an overcomplete basis for the system. Nevertheless, since any real number can be approximated with arbitrary precision by a ratio between two rational numbers, the residual terms in the resolutions of the identity and consequent small corrections can be arbitrarily reduced. 

\subsection{Limit in discretizing time}
As in the previous Section we search here the relativistic limit in discretizing the time values considering $z+1>r_p$ as finite. The spacing between neighboring time values is given again by
\begin{equation}\label{dtau2}
	\delta \tau = \tau_{m + 1} - \tau_{m} = \frac{T}{z+1}
\end{equation}
where here for $T$, from (\ref{en2}), we can derive the following key relation:
\begin{equation}\label{key}
	T = \frac{2\pi\hslash r_n}{E_n} .
\end{equation}
Equation (\ref{key}) must be valid for each $n$ and, in particular for $n=p$, leading to 
\begin{equation}\label{key2}
	T = \frac{2\pi\hslash r_p}{E_p} 
\end{equation}
where the amplitude of the energy spectrum $E_p$ appears explicitly in the denominator.

Combining now equations (\ref{dtau2}) and (\ref{key2}), we obtain:
\begin{equation}\label{39}
	\delta \tau = \frac{2\pi\hslash r_p}{E_p} \frac{1}{z+1} .
\end{equation}
The fundamental inequality is obtained again by requiring that (half) the diameter of the clock $l_C$ not be smaller than its Schwarzschild radius, namely
\begin{equation}\label{39cheserve}
	E_p + m_{\text{rest}}c^2 < \frac{l_C}{2} \frac{c^4}{2G} ,
\end{equation}
which, together with (\ref{39}), leads to 
\begin{equation}\label{finaledisc}
	\delta \tau > 2\pi \left(\frac{l_C}{4l_p t_p} - \frac{m_{\text{rest}}c^2}{\hslash} \right)^{-1} \frac{r_p}{z+1} .
\end{equation}
Given the number of time states $z+1$, equation (\ref{finaledisc}) shows that the limit in the discretization of time values, depending on $m_{\text{rest}}$ and the physical size of the clock. As in the previous Section, this bound on $\delta \tau$ can be made arbitrarily small by choosing $z+1 > r_p$ arbitrarily large. Furthermore, also in this case, under the condition $T > 2\pi r_p/ \left(\frac{l_C}{4l_p t_p} - \frac{m_{\text{rest}}c^2}{\hslash} \right)$ for the period $T$ (which is always satisfied in any realistic scenario), equations (\ref{dtau2}) and (\ref{finaledisc}) ensure that we can safely take the limit $z \rightarrow \infty$, where time values become continuous. Indeed, while $\delta \tau$ tends to zero, the constraint on $\delta \tau$ also tends to zero (with the same scaling), and $\delta \tau$ always remains above the bound.

\subsection{Limit in resolving time}
We briefly discuss here the limit in clock accuracy, working again in the limiting case $z\rightarrow\infty$. The main part of the discussion is essentially the same as that already covered in paragraph II.C and therefore we will not repeat it. The only difference with respect to the previous Section is that, in this case of the generalized quantum clock, we no longer have a set of orthogonal time states. Thus we resort to the \textit{quantum speed limit} \cite{margolus,speedlimit,bush,jan2} to estimate the interval $\Delta t_{\bot} $ required for the state (\ref{timestatecont2}) to evolve into an orthogonal configuration. The structural bound on clock accuracy is thus \cite{speedlimit}:
\begin{equation}\label{4dalpha}
\Delta t \ge	\Delta t_{\bot}  \geq  \text{max} \left( \frac{\pi\hslash}{2 \bar{E}}  , \frac{\pi\hslash}{2 \Delta E} \right) 
\end{equation}
where $\bar{E} = \langle \hat{H} \rangle$ and $\Delta E$ is the spread in energy of the clock given by $\Delta E = \sqrt{\langle (\hat{H} - \bar{E})^2\rangle}$. 

Notice that, by considering $\bar{E}, 2 \Delta E \le E_p$, together with (\ref{39cheserve}) and (\ref{4dalpha}), we can find the (more restrictive of the two) lower limit for $ \Delta t_{\bot}$, namely: 
\begin{equation}\label{ultimo}
	\Delta t_{\bot}  > \pi \left(\frac{l_C}{4l_p t_p} - \frac{m_{\text{rest}}c^2}{\hslash} \right)^{-1}
\end{equation}
which is consistent with what we found in the previous Section in equation (\ref{11}).


As mentioned, when considering the relativistic SW framework, the second bound on $\Delta t$ can be directly obtained by applying the discussion developed in the previous Section, leading to
\begin{equation}\label{limitealtri2}
	\Delta t > 2^{\frac{1}{3}} \Theta^{\frac{1}{3}}t^{\frac{2}{3}}_p 
\end{equation}
which turns out to be a general limit, independent of the structure of the clock energy spectrum. 

The discussion of Section II.C on which bound effectively constrains the clock accuracy for realistic parameter regimes still holds here.
Importantly, the fact that in the generic case the clock period $T$ is no longer rigidly tied to the uniform energy level spacing, but rather is determined through equation (\ref{defT2}), represents a major advantage in two respects. On the one hand, very small energy gaps can make $T$ significantly larger, thereby allowing for longer operational times $\Theta\le T$; on the other hand, a generic spectrum naturally broadens the range of physical systems that can effectively function as quantum clocks. This makes the generic case particularly relevant for realistic implementations.


\section{Conclusions}
 In conclusion, we studied the relativistic limits in discretizing and resolving the time values of a quantum clock, originally introduced in \cite{pegg} and then further developed in \cite{chapter3}. Our clock is represented by an observable complement of a bounded and discrete clock Hamiltonian, which can have an equally-spaced or a generic spectrum: we addressed both cases. We emphasize again that the choice of a bounded Hamiltonian seems the most natural considering that, when we deal with quantum systems, we are always working with systems of finite dimension and the introduction of unbounded Hamiltonians with continuous spectra would not be possible.
 
In the case of clock Hamiltonian with equally-spaced energy spectrum, the (discrete) time observable can be described both by a Hermitian operator (when $z=p$) or by a POVM (when $z>p$). Continuous values for the time observable can be recovered when $z\rightarrow\infty$. 

For $z=p$ we found that the minimum time quantum $\delta \tau_{z=p}$ is actually limited. Nevertheless, we have seen that this limit can be arbitrarily reduced by taking an arbitrarily large $z+1>p$ and we showed that the bound on the minimum time quantum tends to zero in the limit $z\rightarrow\infty$. Thus, we indicated the conditions under which to safely take a continuous flow of time.
When considering the minimum resolvable interval between time values, we derived two bounds through inequalities (\ref{15}) and (\ref{limitealtri}), where the latter is in agreement with previous analysis performed to search for a limit in the accuracy of a time measurement \cite{gambinilimit,limit1,limit2,limit3,limit4}.
As discussed, in realistic parameter regimes the first bound (denoted as \emph{structural}) largely dominates over the relativistic one and must, if anything, be compared with the quantum--spreading constraint (\ref{xmin}) (which effectively becomes the second bound) in determining the actual clock accuracy.
Equation (\ref{limitealtri}) should thus be regarded as an ultimate, fundamental bound that cannot be surpassed, even though realizable implementations remain far from it.


Finally we discussed the generalization of our framework to the case of clock with generic spectrum. In this new scenario we cannot find a subset of $p+1$ time states that are orthogonal, meaning that 
the time observable can only be described by the POVM. Nevertheless, also in this case we found the limit in the discretization of the time values when they are discrete and we show again how to safely take the limit $z\rightarrow\infty$ leading to a continuous flow of time.  
Then we addressed the question of the minimal resolvable interval achievable by the generalized quantum clock. 
We stress once more that, in the generic case, the period $T$ is determined by the energy spectrum through equation (\ref{defT2}), rather than by a fixed uniform spacing. 
This feature offers a twofold advantage: it allows for longer operational times whenever very small energy gaps are present, and it greatly enlarges the class of physical systems that can be employed as clocks. 

However, it is important to emphasize that this work does not aim to propose a realizable clock, but rather to identify fundamental limits on 
the structure and resolution of time values within a specific proposal of a quantum clock \cite{pegg,chapter3,opfase}, when it is constrained by the interplay between quantum mechanics and general relativity. The derived bounds are not engineering prescriptions, but theoretical constraints that define the boundary of what is physically consistent when both quantum uncertainty and relativistic effects are taken into account.

Our original contribution can thus be summarized as follows.  
Assuming a bounded and discrete clock Hamiltonian, for the case in which the time observable is represented by a Hermitian operator we derived a gravitational consistency bound on the discretization of the clock values, identifying a minimum possible time quantum. In the more general case, where the time observable is described by a POVM (which also includes clocks with generic discrete spectra) we determined the condition under which the clock values can be safely treated as continuous, ensuring that this limiting procedure does not violate any constraint imposed by general relativity. Finally, we established a fundamental limit on the resolution of a quantum clock by combining quantum spreading and gravitational arguments, and  explicitly verifying that the Compton bound does not play a role in determining this ultimate limit.  
Importantly, this result is completely general, as it does not depend on the specific structure of the clock spectrum.  
Our discussion thus demonstrates that the fundamental bound applies not only to time measurements, as considered in previous literature, 
but also to the accuracy achievable in distinguishing the values of quantum time operators, independently of the spectral properties of the clock.

As already mentioned, the time observable we studied in this work finds a suitable physical justification (and is thus widely used) within the Page and Wootters quantum time formalism \cite{pagewootters,wootters}. 
In such framework, time is a quantum degree of freedom which belongs to an ancillary Hilbert space and, in such space, it is represented by a clock observable. The Page and Wootters theory is thus a protocol for internalizing the temporal reference frame, leading to a new conception of quantum time where the time observable plays a central role. In the quantum gravity literature, it has been suggested that quantum reference frames are needed in order to formulate a workable quantum theory of gravity \cite{dewitt,afundamental,QG1,QG2}. We hope that our discussion will be useful in this regard but we do not go further into the topic, since it is beyond the scope of the present work.


\section*{Acknowledgements}
The author thanks the Project \lq\lq National Quantum Science and Technology Institute – NQSTI\rq\rq\:Spoke 3 \lq\lq Atomic, molecular platform for quantum technologies\rq\rq. The author also thanks A. Trombettoni for discussions.



\end{document}